\documentclass[draft]{agujournal2019}
\usepackage{url} 
\usepackage{lineno}
\usepackage[inline]{trackchanges} 
\usepackage{soul}
\draftfalse

\journalname{Enter journal name here}

\begin{document}

%
%


\title{Interplanetary magnetic field $B_y$ controlled Alfv\'{e}n wings at Earth during encounter of a coronal mass ejection}

%
%




\authors{Yuxi Chen\affil{1},
Chuanfei Dong \affil{1}, 
Li-Jen Chen\affil{2},
Menelaos Sarantos\affil{2},
Brandon L. Burkholder\affil{2}
}

\affiliation{1}{Center for Space Physics and Department of Astronomy, Boston University, Boston, MA, USA}

\affiliation{2}{NASA Goddard Space Flight Center, Greenbelt, MD, USA }





\correspondingauthor{Yuxi Chen}{yuxichen@bu.edu}




\begin{keypoints}
\item Alfv\'{e}n wings formed at Earth's magnetosphere on 24 April 2023 at the ejecta phase of an interplanetary coronal mass ejection event.
\item Magnetospheric convection patterns driven by dayside and tail reconnection are discussed.
\item Magnetic field, plasma flow, and current system structures around Alfv\'{e}n wings are presented.
\end{keypoints}

%
%

%
%


\begin{abstract}
In the vicinity of Earth's orbit, the typical solar wind Alfv\'{e}n Mach number exceeds 5, and the super-Alfv\'{e}nic solar wind drives a conventional magnetosphere configuration. However, at the ejecta phase of an interplanetary coronal mass ejection (ICME) event, the Alfv\'{e}n Mach number may experience a significant reduction due to the intensified interplanetary magnetic field (IMF) strength and decreased density. On 24 April 2023, an ICME reached Earth’s orbit. The solar wind density dropped to as low as 0.3 amu/cc while the IMF strength is about 25 nT. As a result, the solar wind flow transitions to a sub-Alfv\'{e}nic state with an Alfv\'{e}n Mach number of 0.4, providing opportunities to investigate the interaction of planetary magnetospheres with low Mach number solar wind. We carry out global simulations to investigate the responses of Earth’s magnetosphere to the sub-Alfv\'{e}nic ICME ejecta. The global magnetohydrodynamic (MHD) simulation results show the formation of Alfv\'{e}n wings as the solar wind becomes sub-Alfv\'{e}nic. Furthermore, the sub-Alfv\'{e}nic period was characterized by the dominance of IMF By component, causing the Alfv\'{e}n wings to extend towards the dawn and dusk sides. In this paper, we present the structures of the magnetic field, plasma flow, and current system around the Alfv\'{e}n wings. The global magnetospheric convection under the sub-Alfv\'{e}nic solar wind condition is discussed in depth. Our results achieve a new level of understanding about the interaction between a magnetized body and sub-Alfv\'{e}nic upstream conditions, and provide guidance for future observations.
\end{abstract}

\section*{Plain Language Summary}
In the vicinity of Earth's orbit, the solar wind flow is usually sup-Alfv\'{e}nic, i.e., the flow speed is faster than the local Alfv\'{e}n speed. The interaction between the sup-Alfv\'{e}nic flow and Earth's dipole field generates Earth's bow shock and typical Earth magnetosphere configurations. However, during the ejecta phase of an interplanetary coronal mass ejection (ICME) event, the solar wind density can drop significantly, and the corresponding Alfv\'{e}n speed increases so that it is larger than the flow speed, i.e., the flow becomes sub-Alfv\'{e}nic. The bow shock would disappear when the upstream solar wind is sub-Alv\'{e}nic, and Earth's magnetosphere reconfigures to a new state that is characterized by two Alfv\'{e}n wings. On 24 April 2023, the solar wind became sub-Alfv\'{e}nic during an ICME event. We simulated this event with a global magnetohydrodynamics (MHD) model. This paper presents the structures of the plasma flow and the magnetic field lines around the Alfv\'{e}n wings and discusses the global convection patterns.


%
%
\section{Introduction}
The solar wind is usually super-Alfv\'{e}nic around Earth's orbit. The interaction between the super-Alfv\'{e}nic solar wind and Earth's intrinsic dipole magnetic field forms the bow shock in front of Earth's magnetosphere. In the direction that is perpendicular to the Sun-Earth line, Earth's magnetosphere is usually confined within a radius of 20-30 Earth radius ($R_E$) by the bow shock. Occasionally, the solar wind flow can become sub-Alfv\'{e}nic due to the increase of the Alfv\'{e}n speed, especially during an interplanetary coronal mass ejection (ICME) event when the interplanetary magnetic field (IMF) is strong and the plasma density becomes low. In the sub-Alfv\'{e}nic solar wind, the bow shock disappears, and the Alfv\'{e}n wave originating near Earth can propagate upstream freely without being confined by the shock. As a consequence, the open field lines can extend a few hundred $R_E$ away from the Earth in the direction that is perpendicular to the solar wind flow and forms the structure called Alfv\'{e}n wings. 

Alfv\'{e}n wings were originally found by studying the interaction between a satellite and the background plasma flow \cite{Drell:1965}. Since several moons, such as Ganymede \cite{Kivelson:1998} and Io \cite{Goertz:1980}, are exposed to sub-Alfv\'{e}nic flows most of the time, Alfv\'{e}n wings have also been found there. Alfv\'{e}n wings are featured by open magnetic field lines that are connected to the obstacle body and extend outward. The angle between these open field lines and the incident IMF is determined by the ratio between the solar wind speed and the Alfv\'{e}n speed, i.e., $\theta = atan(V_{sw}/V_{A})$. The open field line tubes are obstacles for the solar wind and divert the flow. Inside the tubes, the plasma flow is much slower than the surrounding solar wind flow. 

The structures of Alfv\'{e}n wings have been investigated with both satellite data and numerical simulations, especially for the moons \cite{Kopp:2002, Ip:2002, Jia:2008,wang2018electron,zhou2019embedded}. Since the sub-Alfv\'{e}nic solar wind is rare near Earth's orbit, only a few events have been observed in the past decades \cite{usmanov2005low}, and fewer events have been investigated with global simulations. The Alfv\'{e}n wings at Earth's magnetosphere are usually observed during the ejecta phase of an ICME event when the solar wind density is low and Alfv\'{e}n speed is large \cite{lugaz2016earth,chane2012observational}. \citeA{ridley2007alfven} performed a series of simulations to investigate the Alfv\'{e}n wing structures and their impacts on global geoeffects \cite{kivelson2008saturation}. \citeA{chane2012observational} and \citeA{chane2015simulations} studied an event with both satellite data and global simulations and investigated the magnetic field and plasma flow signatures of the Alfv\'{e}n wings.

On 23 and 24 April 2023, an ICME event arrived at Earth's orbit. We simulate the whole event with the global magnetohydrodynamic (MHD) model BATS-R-US \cite{Powell:1999}, and find Alfv\'{e}n wings formed around 13:00:00 UT on 24 April 2023 when the solar wind is sub-Alfv\'{e}nic. In this paper, we present the comparison between the MMS observations and the simulation for validation in Section~\ref{section:mms1_vs_sim}. The magnetic field and plasma flow structures are presented in Section~\ref{section:field-and-flow}, and the magnetospheric convection patterns are discussed in Section~\ref{section:convection}. The current system is illustrated in Section~\ref{section:current}. Finally, we summarize the results in Section~\ref{section:summary}.

\section{Event Overview and Simulation Setup}

\begin{figure}
  \includegraphics[width=1.0\textwidth]{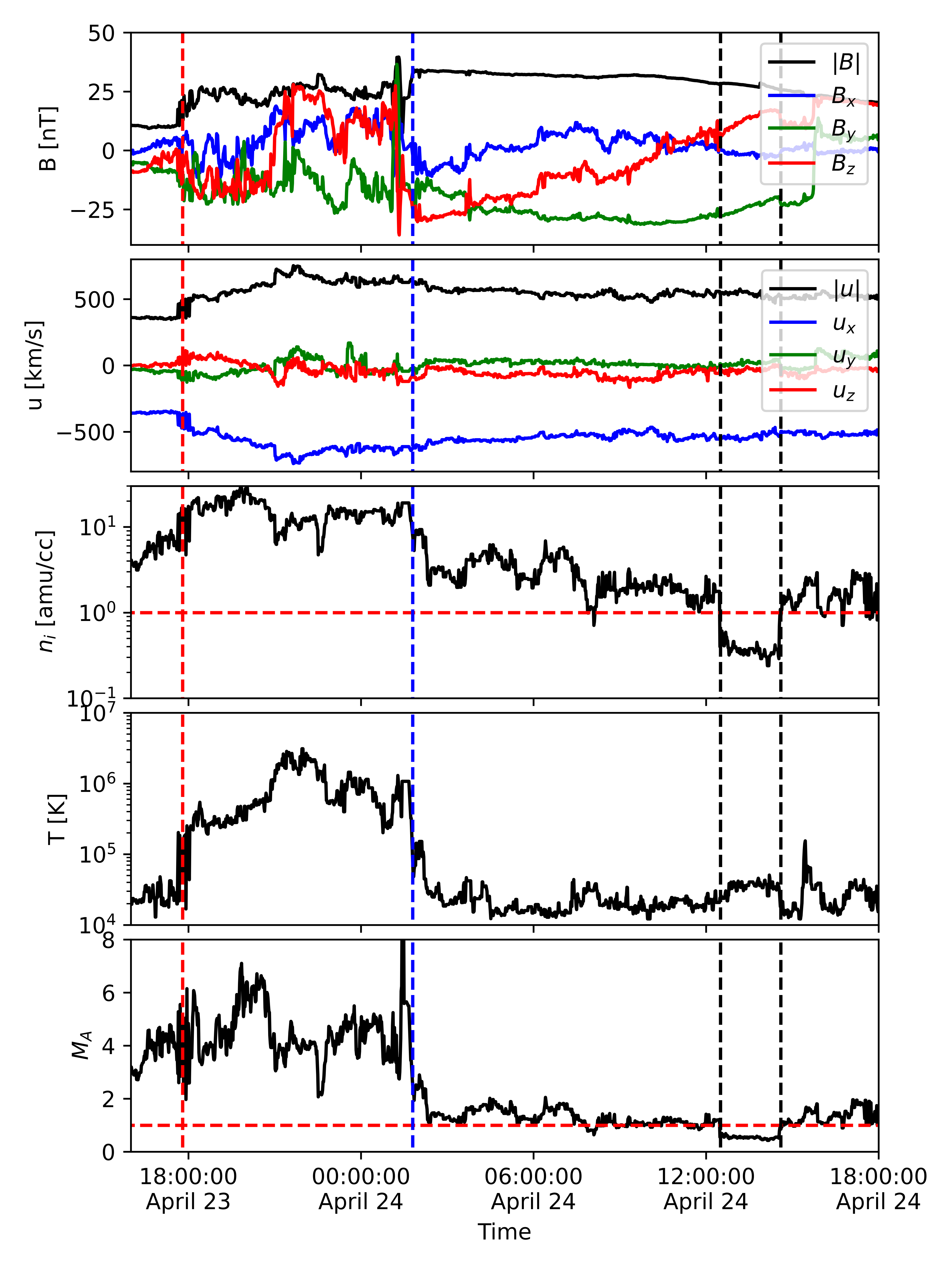}
  \caption{Solar wind data from the Wind spacecraft. The data has been shifted to the bow shock nose. The vertical red line indicates the arrival of the ICME shock. The vertical blue line indicates the start of the ICME ejecta phase. The two vertical black lines show the period when the solar wind Alfv\'{e}n Mach number is well below 1.}
  \label{fig:wind}
\end{figure}

Figure~\ref{fig:wind} shows the solar wind data from the Wind spacecraft for the ICME event on 23 and 24 April 2023. We note that Wind is located at the L1 point, but the data shown in Figure~\ref{fig:wind} has been shifted to the bow shock nose. The shock of the ICME arrives at Earth around 18 UT on 23 April 2023 (vertical red line in Figure~\ref{fig:wind}), and the ICME sheath region is observed in the following 8 hours (the time period between the red and blue vertical lines in Figure~\ref{fig:wind}). Around 2 UT on 24 April 2023 (blue vertical line in Figure~\ref{fig:wind}), the ICME ejecta phase starts and the IMF $B_z$ component drops from about 20 nT to -30 nT abruptly. Subsequently, the IMF $B_z$ takes about 14 hours to gradually increase to 20 nT, which corresponds to the geomagnetic storm recovery phase. At around 12:30:00 UT on 24 April 2023, the solar wind density drops to as low as 0.3 amu/cc and lasts for about two hours, while the IMF strength is still about 25 nT. As a result, the solar wind flow becomes sub-Alfv\'{e}nic with an Alfv\'{e}n Mach number as low as 0.4 (the period between the two vertical black lines in Figure~\ref{fig:wind}). In the sub-Alfv\'{e}nic solar wind, the magnetic perturbation produced at Earth's magnetopause can propagate upstream without bound, and the bow shock is expected to expand away from the Earth and eventually vanish. Earth's magnetosphere reconfigures to a new state, which is characterized by two Alfv\'{e}n wings.

As demonstrated below, MMS crossed the magnetopause at about 14:24:00 UT on 24 April 2023 when Earth's magnetosphere was exposed to the sub-Alfv\'{e}nic solar wind. It is difficult to reveal the global configurations of the Alfv\'{e}n wings with satellite data alone. Therefore, we carry out global MHD simulations with the well-established BATS-R-US model to investigate Alfv\'{e}n wings' structures. Since the solar wind can become sub-Alfv\'{e}nic and the bow shock expands away from the Earth, the simulation domain is much larger than typical simulations to ensure the upstream boundary does not interfere with the magnetosphere. To be specific, the simulation domain is $-512~R_E < x < 256~R_E$ and $-256~R_E < y,z < 256~R_E$ in the GSM coordinates. Upstream of the Alfv\'{e}n wings, the grid resolution is at least $\Delta x = 1~R_E$, and it is at least $\Delta x = 0.5~R_E$ around the Alfv\'{e}n wings. Near the dayside magnetopause and inner magnetosphere, the grid resolution can be as high as $\Delta x = 0.125~R_E$. We refer readers to the simulation parameter file, which can be obtained from the link provided in the Acknowledgments section, for more details. The magnetosphere model is coupled to a two-dimensional ionospheric model \cite{Ridley:2004}, which provides the velocity boundary conditions for BATS-R-US, through the Space Weather Modeling Framework (SWMF) \cite{Toth:2005swmf,Toth:2012swmf}.

The simulation starts from 16 UT on 23 April 2023 and lasts 26 hours, but we only focus on the period when the solar wind is sub-Alfv\'{e}nic in this paper. We note that during the sub-Alfv\'{e}nic period, the IMF is dominated by the $B_y$ component, and the $B_z$ component is small and northward, which is different from the majority of the previous simulations for Alfv\'{e}n wings, including the simulations of Earth's Alfv\'{e}n wings performed by \citeA{ridley2007alfven}.

\section{MMS Observation-Simulation Comparison}
\label{section:mms1_vs_sim}
\begin{figure}
  \includegraphics[width=1.0\textwidth]{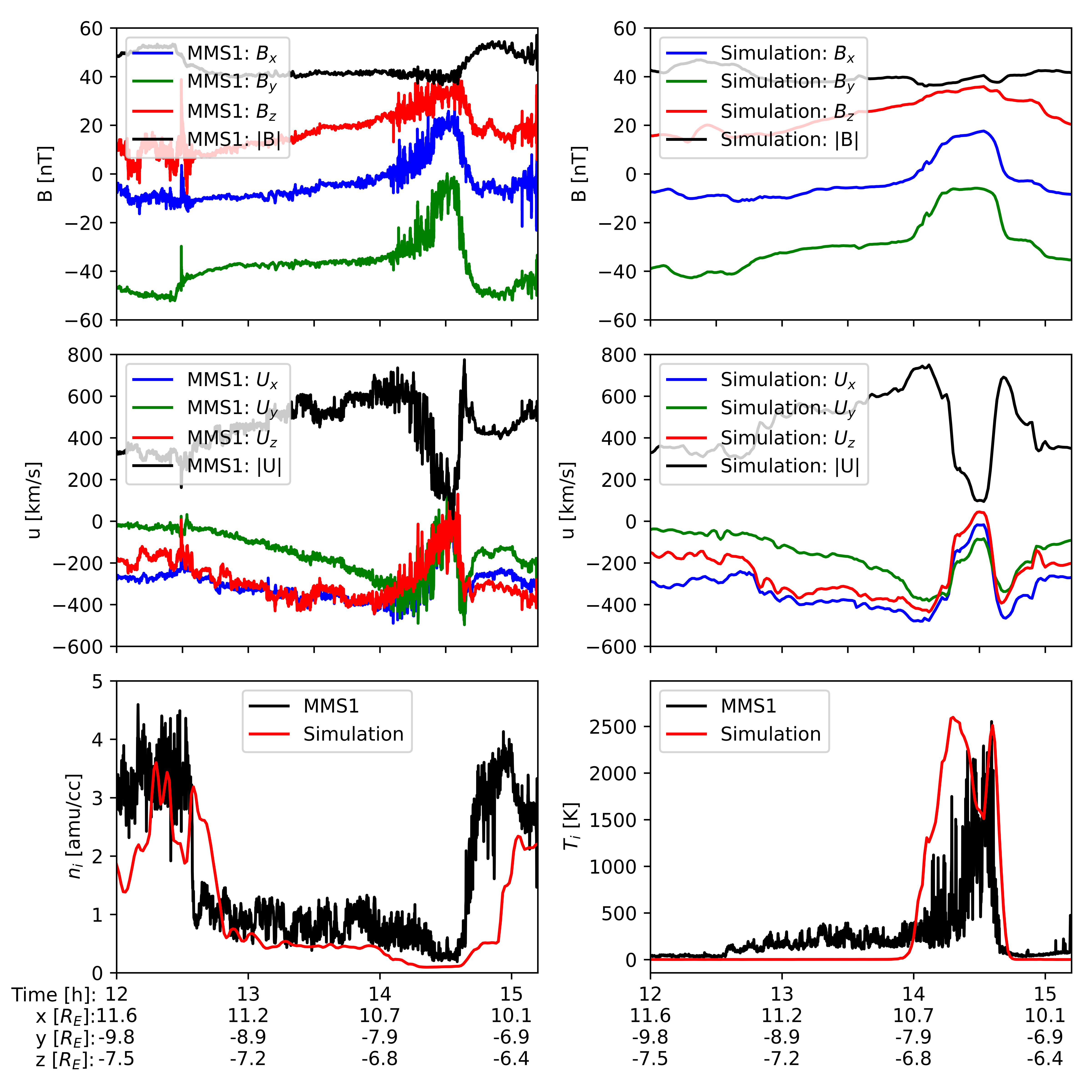}
  \caption{MMS1 data and the simulation results along a virtual satellite, which is close to MMS1's trajectory. The x-axis labels are the time and MMS1 locations in GSE coordinates. The simulated magnetic field and velocity are converted from GSM to GSE for comparison.}
  \label{fig:mms1_vs_sim}
\end{figure}

Figure~\ref{fig:mms1_vs_sim} displays the magnetic fields and plasma quantities obtained from both MMS1 observations and the global MHD simulation results. The simulation data was acquired along a virtual satellite orbit that is close to MMS1's trajectory with time-varying simulation outputs. In Figure~\ref{fig:field-line-flow}(a) and (c), the black arrows represent the virtual satellite's trajectory relative to the magnetosphere structures. At approximately 12:40:00 UT, MMS1 detected a sudden density reduction from 4 amu/cc to below 1 amu/cc, consistent with the density drop observed in the Wind data around 12:30:00 UT. The comparable low-density values observed by both Wind and MMS1 suggest bow shock is either absent or extremely weak between these spacecraft during this period. Subsequently, MMS1 navigated toward the magnetopause, entering the magnetosphere around 14:20:00 UT. Roughly 10 minutes later, due to the motion of the magnetopause caused by increased solar wind density and dynamic pressure, MMS1 transitioned back to the solar wind. The simulation results mirror the trends in MMS1 data across all parameters, validating the simulation's fidelity in capturing the magnetosphere configuration during the sub-Alfv\'{e}nic period.

\section{Magnetic Field and Plasma Flow Structures}
\label{section:field-and-flow}

\begin{figure}
  \includegraphics[width=1.0\textwidth,trim=0cm 4cm 0cm 0cm]{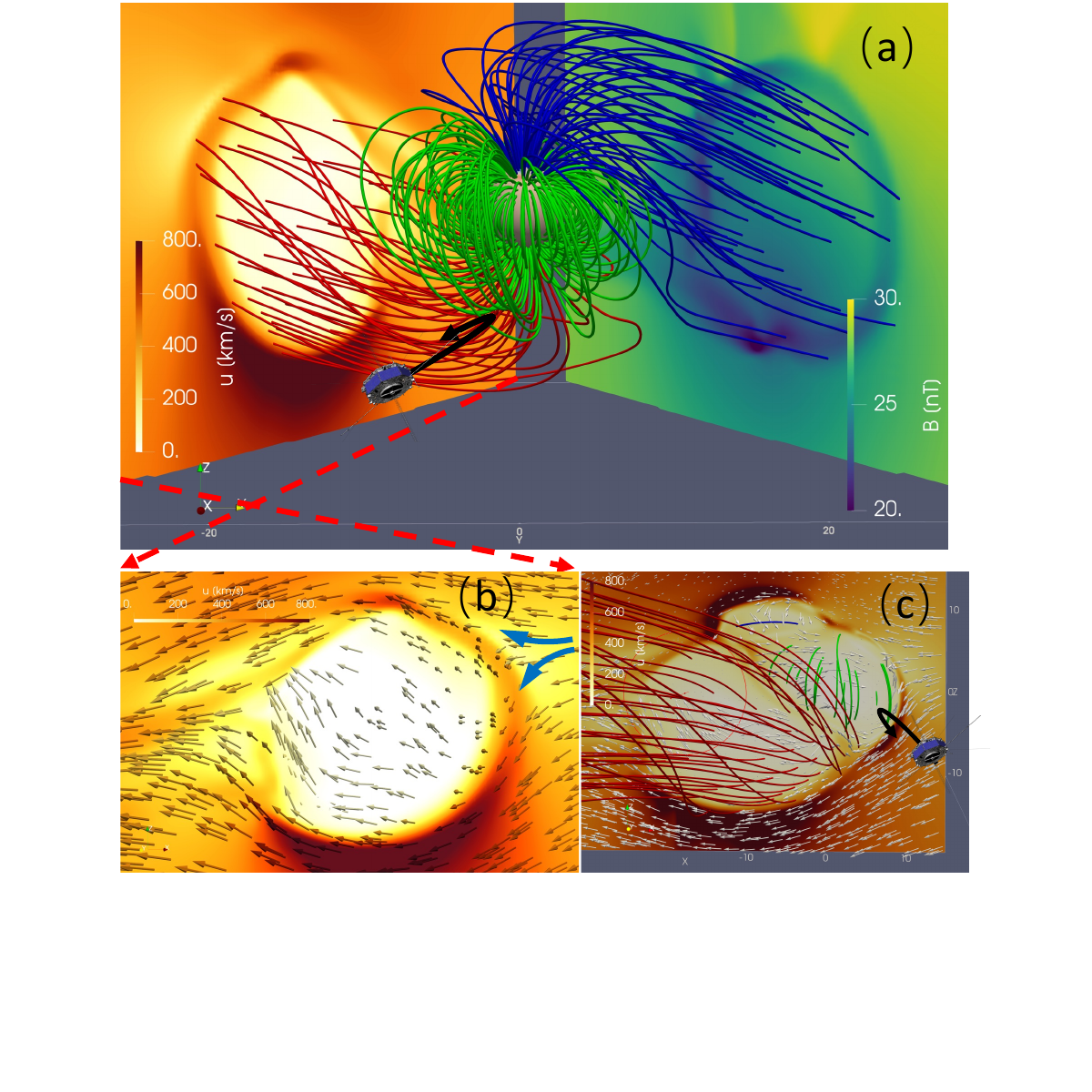}
  \caption{(a) Magnetic field lines and 2D planes cross Alfv\'{e}n wings which show the magnetic field strength (duskside) and the plasma speed (dawnside). (b) The plasma speed and velocity directions (arrows) around the dawnside Alfv\'{e}n wing. Panel(c) shows the same quantities as in (b) but in the plane of $y=-7~R_E$. The black arrows in panels (a) and (c) represent the simulated virtual satellite trajectory relative to the Alfv\'{e}n wing.}
  \label{fig:field-line-flow}
\end{figure}

When the IMF is dominated by the $B_z$ component, the open field lines that constitute the Alfv\'{e}n wings stretch in the north-south direction \cite{ridley2007alfven}. For the event presented here, IMF $B_y$ is the dominant component, so the open field lines stretch along the dawn-dusk direction as shown in Figure~\ref{fig:field-line-flow}(a). These magnetic field lines are traced from the surface of $r=3R_E$ with uniformly distributed seeds. The open field lines emanating from the northern and southern hemispheres are colored blue and red, respectively. The closed field lines are colored with green. All the northern/southern hemisphere open field lines extend towards the dusk/dawn side (+Y/-Y) and form two Alfv\'{e}n wings, which slow down the incident solar wind flow as obstacles. Figure~\ref{fig:field-line-flow}(b) shows the plasma speed and velocity directions in a plane that is perpendicular to the duskside Alfv\'{e}n wing. The plasma inside the wing flows away from the Earth along the magnetic field lines with a speed that is much smaller than the solar wind speed. Since the wings behave like obstacles, the plasma flow is deflected along the surface of the wing. The plasma flux inside a given streamtube is $F = \rho u \Delta A$. Assume that the plasma flow reaches a quasi-steady state around the wing, and its density does not vary too much along a streamline. Due to the deflection caused by the Alfv\'{e}n wings, the cross-section of the streamtube reduces near the surface of the wing, and the plasma flow must be accelerated to conserve the plasma flux. That is why the plasma speed exceeds $800~km/s$ around the wing in Figure~\ref{fig:field-line-flow}. Figure~\ref{fig:field-line-flow}(c) shows the same quantities as in (b) but in the plane of $y=-7~R_E$, which is near the MMS1 trajectory. Panel (c) demonstrates the plasma speed is also large at the edge of the wing. MMS1 enters the closed field line region, and it is consistent with the velocity dips in Figure~\ref{fig:mms1_vs_sim}. 

In both Figure~\ref{fig:field-line-flow}(b) and (c), the speed below the wing is significantly higher than the speed above the wing. As the blue arrows indicate, in front of the dawnside wing, more plasma is deflected to the bottom of the wing than to the top, so the plasm speed is larger below the wing. We note that this asymmetry pattern is oppositive for the duskside wing, as can be seen from Figure~\ref{fig:convection}(c). 

\section{Magnetospheric Convection}
\label{section:convection}

\begin{figure}
  \includegraphics[width=1.0\textwidth,trim=0cm 4cm 0cm 0cm]{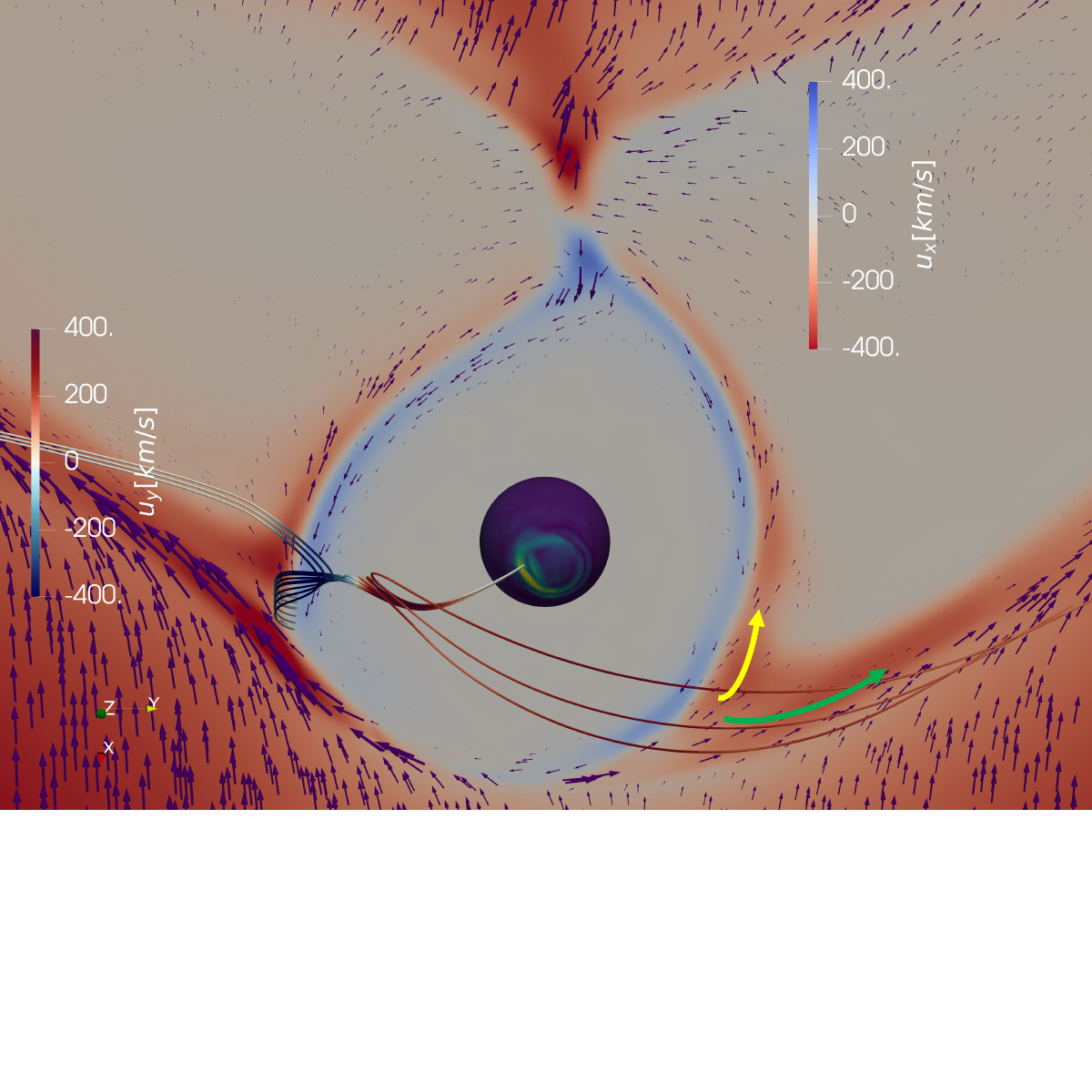}
  \caption{The z=0 plane colored by the plasma $u_x$ component. The magnetic field lines are colored by the $u_y$ component. The small blue arrows represent the flow directions in the x-y plane.}
  \label{fig:xy-ux}
\end{figure}

\begin{figure}
  \includegraphics[width=1.0\textwidth]{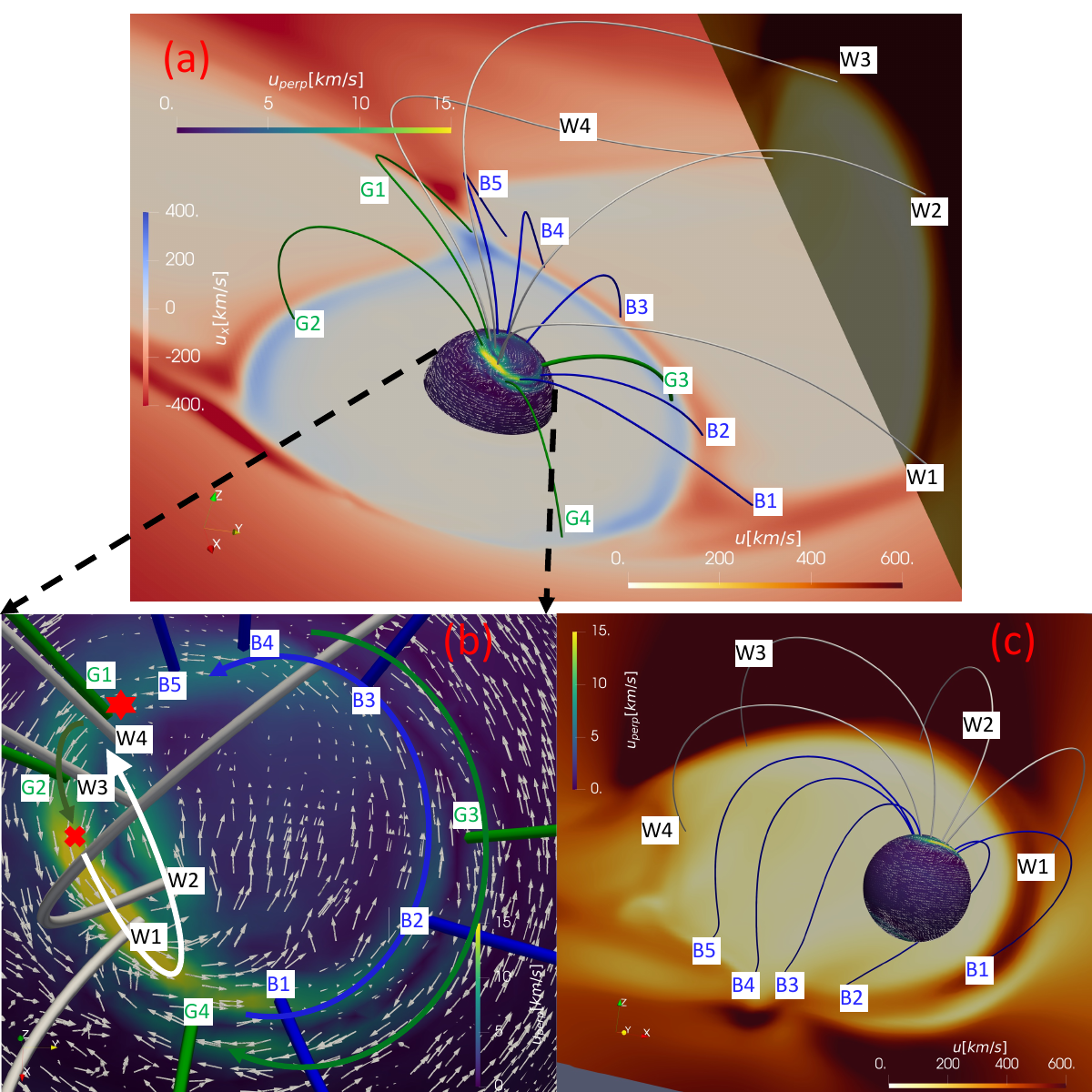}
  \caption{Magnetic field line convection. Panel (a) shows the z=0 plane colored by $u_x$. Panel (a) and (c) contain a plane that cuts through the dusk-side Alfen wing and shows the plasma speed ($u$). Panel (b) shows the magnetic field line convection speed $|u_{perp}|$ at $r=3R_E$, and the arrows represent the directions of the convection. White and blue field lines are open, and green field lines are closed.}
  \label{fig:convection}
\end{figure}

\begin{figure}
  \includegraphics[width=1.0\textwidth, trim=0cm 12cm 0cm 0cm]{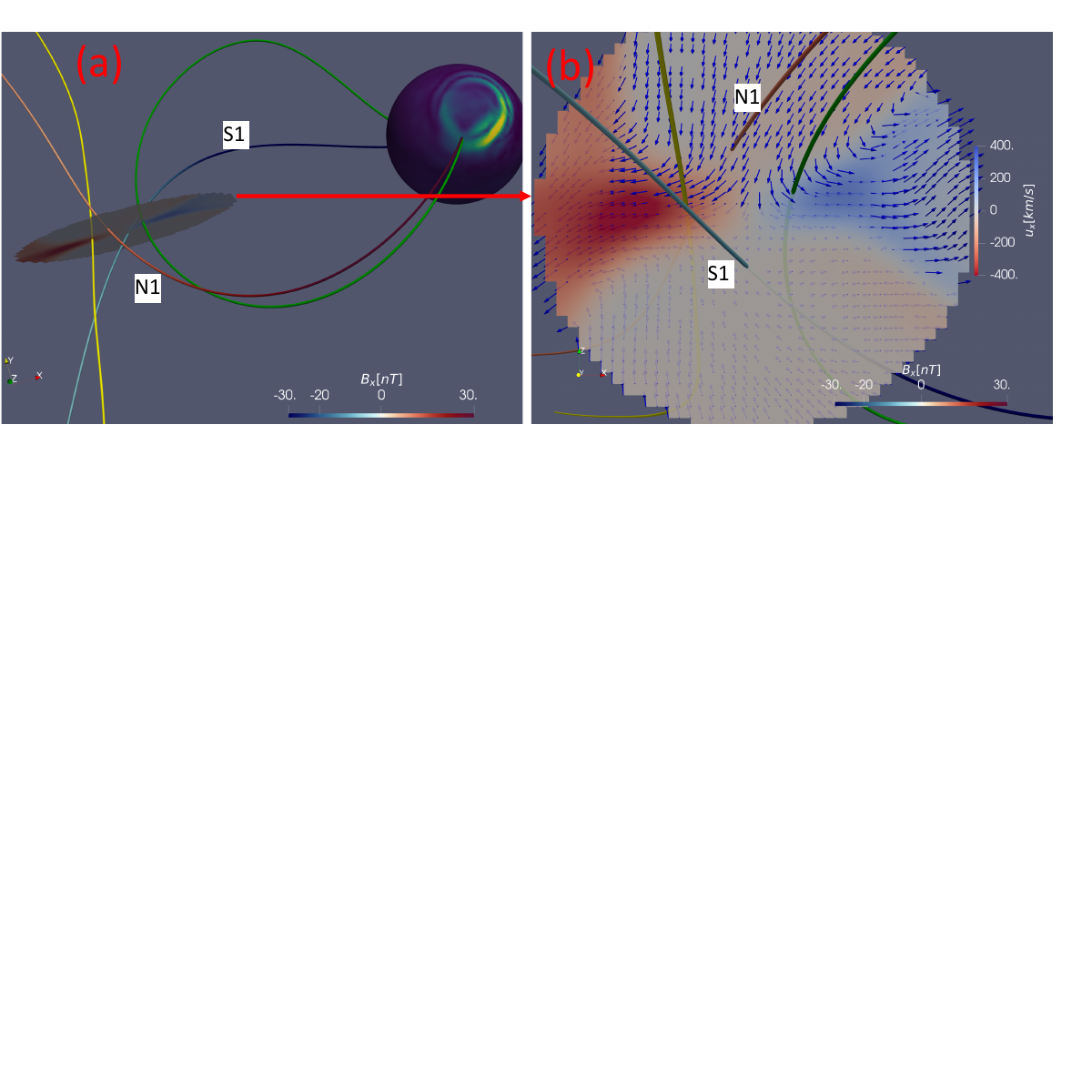}
  \caption{(a) Magnetic field lines near the tail current sheet. Line S1 and N1 are in the southern and northern lobes, respectively, and they are colored by the $B_x$ field. The green line is the reconnected closed field line, and the yellow line is the reconnected field line that is connected to the solar wind on both sides. (b) Plasma $u_x$ component in the y=0 plane and the 3D flow directions (blue arrows).}
  \label{fig:tail}
\end{figure}

Figure~\ref{fig:xy-ux} displays the plasma velocity $u_x$ component in the equator plane. The blue arrows represent the flow directions in the plane. A few magnetic field lines, which are colored by the $u_y$ component, are plotted to show the location of magnetic reconnection in the northern hemisphere. The curvature of the field lines and the reversal of the $u_y$ component are the evidence of magnetic reconnection. Due to the symmetries of the dipole and IMF field, magnetic field lines also reconnect in the southern hemisphere on the duskside (+Y). For simplicity, we focus on the reconnection in the northern hemisphere. After reconnection, the open field lines will be transported to the tail by the plasma flow. Yellow and green arrows in Figure~\ref{fig:xy-ux} represent the two channels for tailward convection. After the open field lines reach the tail, they will reconnect again to close the field lines. The reversal of $u_x$ in the equatorial plane indicates the location where tail reconnection occurs. The positive $u_x$ area is the path for the sunward convection of closed field lines. This path is near the open-closed field line boundary, as it is demonstrated below in Figure~\ref{fig:convection}(a).

Figure~\ref{fig:convection} depicts the convection patterns of magnetic field lines. White lines (W1 to W4) and blue lines (B1 to B5) represent the open field lines of two different convection cells, and the green lines (G1 to G4) are closed field lines. Panel (c) shows the plasma speed in a plane that cuts through the duskside Alfv\'{e}n wing. Panel (b) displays the contour and directions (small white arrows) of the magnetic field convection velocity $u_{perp}$ on the surface of $r=3~R_E$, which is defined as the velocity component that is perpendicular to the local magnetic field. The red cross represents the footpoints of the just reconnected magnetic fields that are plotted in Figure~\ref{fig:xy-ux}. After reconnection, the open field lines are dragged towards the dayside first (from the red cross to w1) due to the release of the magnetic tension force. Afterwards, they are transported tailward (w1-w2-w3-w4) to form the small dawnside convection cell (white arrow). Magnetic reconnection can also happen in the southern hemisphere on the duskside, the footpoints of these open field lines in the northern hemisphere are the blue lines (B1 to B5), and they are moving tailward (B1-B2-B3-B4-B5) to generate the large duskside convection cell (blue arrow). Panel (c) shows the footpoints of the open field lines in the duskside Alfv\'{e}n wing. The dawnside cell open field lines (w1-w2-w3-w4) are transported along the upper edge of the wing, and the duskside cell open field lines (B1-B2-B3-B4-B5) are transported along the lower edge of the wing. Since the blue lines connect the northern hemisphere and the lower edge of the wing, they cross the equatorial plane and form the tailward field line transportation channel along the open-closed field line boundary (yellow arrow in Figure~\ref{fig:xy-ux}). 

After the open field lines reach the tail, they will reconnect again to close the field lines. Figure~\ref{fig:tail} shows a few magnetic field lines before and after reconnection. Field lines S1 and N1 are in the southern and northern lobes, respectively. They are open and colored by the $B_x$ field. Although both S1 and N1 contain a large $B_y$ component, they can still reconnect due to the reversed $B_x$ component. The velocity arrows and the $u_x$ component in the $y=0$ plane (Figure~\ref{fig:tail}(b)) demonstrate that the plasma and magnetic fields flow towards the center, reconnect, and then leave the reconnection region. The closed field lines are transported back to dayside magnetopause along the path of G1-G2 or G1-G3-G4 in Figure~\ref{fig:convection} (green arrows in panel (b)). The paths of G1-G2 and G1-G3-G4 complete the dawnside and duskside convection cells, respectively. We note that these closed field lines are adjacent to the open field line region (for example, B2-B5 lines). It demonstrates the sunward convection channel (G1-G2 or G1-G3-G4), and one of the tailward convection channels (B1-B2-B3-B4-B4 in Figure~\ref{fig:convection} or the yellow arrow in Figure~\ref{fig:xy-ux}) are close to the open-closed field line boundary.

The global magnetospheric convection patterns presented here share some similarities with conventional Dungey convection with strong IMF $B_y$. Figure~10(c) in \citeA{sandholt1998} is a sketch that shows the dayside convection pattern with $Bz\sim 0$ and $B_y<0$ under super-Alfv\'{e}nic solar wind conditions, and it is very similar to the pattern shown in Figure~\ref{fig:convection}(b). This is because the Alfv\'{e}n wings do not significantly change the magnetic field topology of the inner magnetosphere, and the convection pattern in the inner magnetosphere and ionosphere is mainly controlled by the IMF $B_y$ component.

\section{Current System}
\label{section:current}

\begin{figure}
  \includegraphics[width=1.0\textwidth, trim=0cm 12cm 0cm 0cm]{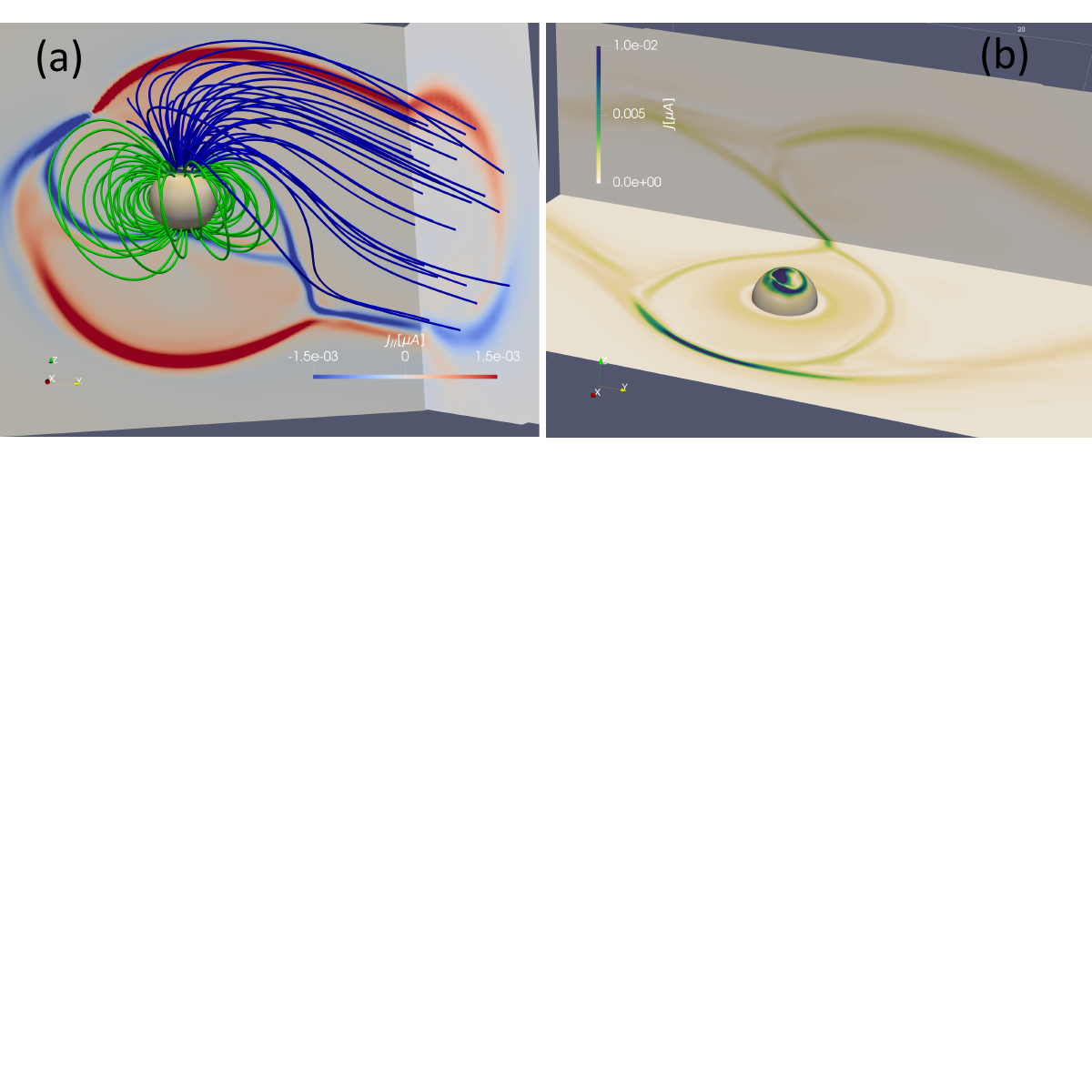}
  \caption{(a)$J_{\parallel}$ in two planes, which is either quai-parallel or quai-perpendicular to the Alfv\'{e}n wing field lines. (b) Current strength $J$ in the planes of $z=0$ and $x=-15R_E$.}
  \label{fig:current}
\end{figure}

At the edge of the Alfv\'{e}n wings, there are strong field-aligned currents (FACs) (Figure~\ref{fig:current}(a)), which is consistent with previous analytical studies and simulations \cite{Drell:1965, Neubauer:1980, ridley2007alfven}. It is well-known that a strong IMF $B_y$ component twists the tail current sheet under super-Alfv\'{e}nic solar wind conditions \cite{kaymaz1994interplanetary}, and Figure~\ref{fig:current}(b) shows that the short tail current sheet in this event is also twisted.


\section{Summary and Conclusions}
\label{section:summary}

In this paper, we present the simulation of the Alfv\'{e}n wings at Earth's magnetosphere during the ejecta phase of the ICME on 24 April 2023. The plasma flow and magnetic field structures around the Alfv\'{e}n wings are examined, and the magnetospheric convection patterns are discussed in detail. The reconnected magnetic field lines on the dayside are transported to magnetotail along the edge of the Alfv\'{e}n wings, and the open field lines reconnect again in the tail to produce closed field lines. The closed field lines flow to the dayside along the open-closed field line boundaries to complete a convection cycle. The concept is similar to the Dungey cycle under super-Alfv\'{e}nic solar wind conditions, but the field line paths far away from the inner magnetosphere are mediated by the Alfv\'{e}n wings. This study facilitates our understanding of the interaction between a magnetized body and sub-Alfv\'{e}nic upstream conditions, and provides guidance for future observations.


\section{Open Research}
Wind spacecraft data were obtained from the CDAWeb (https://cdaweb.gsfc.nasa.gov/index.html/). MMS data are publicly available at https://lasp.colorado.edu/mms/sdc/public/. The SWMF model, including BATS-R-US and the ionosphere model, is publicly available through GitHub (https://github.com/SWMFsoftware). The simulation input files and output are available through the Zenodo repository (https://doi.org/10.5281/zenodo.10445643).

\section{Conflict of Interest}
The authors declare no conflicts of interest relevant to this study.

\acknowledgments
Computational resources supporting this work were provided by the NASA High-End Computing (HEC) Program through the NASA Advanced Supercomputing (NAS) Division at Ames Research Center. This work is supported by NASA under grants 80NSSC22K0323 and 80NSSC24K0144, NSF grant AGS-2149787, and NASA's MMS Mission.


%
%



\bibliography{csem,alfven_wing}

%
%
%
%
%

\end{document}